\documentclass[runningheads,a4paper]{llncs}

\usepackage{amssymb}
\usepackage{graphicx}

\usepackage{multirow}

\usepackage{url}
\urldef{\mailsa}\path|econst@csd.auth.gr|
\urldef{\mailsb}\path|gkakaron@teilar.gr|
\urldef{\mailsc}\path|stamelos@csd.auth.gr|
\newcommand{\keywords}[1]{\par\addvspace\baselineskip
\noindent\keywordname\enspace\ignorespaces#1}
\begin{document}

\mainmatter  

\title{Open Source Software: How Can Design Metrics Facilitate Architecture Recovery?}

\titlerunning{How Can Design Metrics Facilitate Architecture Recovery?}

\author{Eleni Constantinou\inst{1} \and George Kakarontzas\inst{2} \and Ioannis Stamelos\inst{1}}
%

\institute{Computer Science Department\\
Aristotle University of Thessaloniki\\
Thessaloniki, Greece \\
\mailsa\\
\mailsc\\
\and
Computer Science Department\\
Aristotle University of Thessaloniki\\
Thessaloniki, Greece, and\\
Computer Science and Telecom. Department\\
T.E.I. of Larisa, Greece\\
\mailsb\\
}
%
%
\maketitle

\begin{abstract}
Modern software development methodologies include reuse of open source code. Reuse can be facilitated by architectural knowledge of the software, not necessarily provided in the documentation of open source software. The effort required to comprehend the system's source code and discover its architecture can be considered a major drawback in reuse. In a recent study we examined the correlations between design metrics and classes' architecture layer. In this paper, we apply our methodology in more open source projects to verify the applicability of our method.
\end{abstract}

\section{Introduction}
\label{sec_introduction}
Software Architecture is considered \emph{`the set of structures needed to reason about the system, which comprise software elements, relations among them, and properties of both'} \cite{Clements2010}. However, open source software documentation does not always facilitate system understanding. Consequently, developers have to inspect the system's source code in order to reuse it. Since source code is available, design metrics can be used for architecture recovery of open source systems \cite{Buschmann1996}.

While most architecture recovery techniques mainly focus on clustering and pattern-based techniques, our approach is based on static source code analysis \cite{Constantinou2011}. We use the Directed Acyclic Graph(DAG) of the system, produced from static analysis, and derive D-layers (DAG Layers). D-Layers are subsets of classes, where subsets at higher D-Layer depend only on the subset of classes in lower D-layer and not vice versa. Although architectural layers cannot reveal classes' role in the system (user interface, domain class, etc.), they can provide a starting point for the actual layer recovery. Since there is significant statistical variations in D-Layer metrics, we use design metrics to predict D-layers and form \emph{tentative} architectural layers. D-layer prediction is accomplished using machine learning techniques, and more specifically classification rules generation relating metric values to architecture layers.

\section{Methodology}
\label{sec_methodology}
Software systems following layered architecture are usually consisted of four layers: \emph{User Interface} (4), \emph{Controllers} (3), \emph{Business Logic} (2) and \emph{Infrastructure} (1). Our initial hypothesis is that classes in different layers have different functional behavior and dependencies with classes in other layers. User Interface classes are focused in information presentation and interaction with system users, while they obtain information from classes in Controller layer. Business Logic layer is consisted of domain-specific classes that use Infrastructure Layer classes for data storage or other general-purpose services.

Our methodology steps are:
\begin{enumerate}
   \item Form the DAG of the system using static analysis to derive D-layers.
   \item Arbitrarily combine consecutive D-layers to form four groups of \emph{tentative} architectural layers
   \item Calculate the Chidamber and Kemerer metrics for each class of the system.
   \item Calculate the correlation of the indicative metrics of architectural layers to the \emph{current architectural partition}
   \item Extract classification rules that utilize design metrics to predict layers.
 \end{enumerate}

Chidamber and Kemerer \cite{Chidamber1999} metrics measure class-level properties and are calculated by Chidamber and Kemerer Java Metrics (ckjm) software package \cite{Spinellis2005}. Our methodology includes Weighted Methods per Class (WMC), Depth of Inheritance Tree (DIT), Number of Children (NOC), Coupling Between Object (CBO), Response For a Class (RFC), Lack of Cohesion in Methods (LCOM), Afferent Coupling (Ca) and Number of Public Methods (NPM).

The validation of our methodology for each architecture layer is based on the following measures:
\newline \emph{Precision}: the number of correctly classified instances, divided by the total number of instances classified in this layer.
\newline \emph{Recall}: the number of correctly classified instances, divided by the total number of instances of this layer.

\section{Results}
\label{sec_results}
Next, we provide a description of the data used for the validation of the proposed method. We apply our method to four large open source systems written in Java, JabRef, Jbpm, RapidMiner and SweetHome3D.  JabRef \cite{JabRef} is an open source bibliography reference manager which consists of 3265 classes, Jbpm \cite{Jbpm} is a flexible business project management suite consisted of 627 classes, RapidMiner \cite{RapidMiner} is a data analysis and data mining system consisted of 4389 classes and SweetHome3D \cite{SweetHome3D} is a free interior design application consisted of 1284 classes.

Initially, we extract classes' D-layers using Classycle \cite{Classycle} and the corresponding design metric values using ckjm. Some descriptive statistics are shown for JabRef, Jbpm, RapidMiner and SweetHome3D in Tables \ref{table_metrics_JabRef}- \ref{table_metrics_SweetHome3D}.

\begin{table}[!t]
\renewcommand{\arraystretch}{1.3}
\caption{Descriptive Statistics for JabRef}
\label{table_metrics_JabRef}
\centering
\begin{tabular}{|c|c|c|c|c|c|c|c|c|c|c|}
\hline
 & Min & Max & Mean & Std. Deviation & Bin=1 & Bin=2 & Bin=3 & Bin=4 & Bin=5 & Bin=6\\
\hline
D-layer & 0 & 15 & 6.57 & 4.761 & 0-3 & 4-7 & 8-11 & 12-15 &  &  \\
WMC & 0 & 309 & 7.96 & 14.393 & 0 & 1 & 2 & 3-28 & 29-309  & \\
DIT & 1 & 8 & 1.54 & 1.217 & 0-5 & 6-8 &  & &  & \\
NOC & 0 & 41 & .27 & 1.728 & 0 & 1-41 &  & &  & \\
CBO & 0 & 111 & 4.60 & 6.419 & 0 & 1 & 2-7 & 8-111 &  & \\
RFC & 0 & 450 & 23.18 & 35.005 & 0 & 1-3 & 4 & 5-10 & 11-42 & 43-450 \\
LCOM & 0 & 46616 & 83.33 & 1074.946 & 0 & 1-46616 &  & &  & \\
Ca & 0 & 441 & 4.50 & 15.502 & 0 & 1 & 2-441 & &  & \\
NPM & 0 & 298 & 5.83 & 11.275 & 0 & 1 & 2 & 3-7 & 8-298 & \\
\hline
\end{tabular}
\end{table}

\begin{table}[!t]
\renewcommand{\arraystretch}{1.3}
\caption{Descriptive Statistics for Jbpm}
\label{table_metrics_Jbpm}
\centering
\begin{tabular}{|c|c|c|c|c|c|c|c|c|}
\hline
 & Min & Max & Mean & Std. Deviation & Bin=1 & Bin=2 & Bin=3 & Bin=4\\
\hline
D-layer & 0 & 19 & 4.82 & 4.034 & 0-4 & 5-9 & 10-14 & 15-19 \\
WMC & 0 & 70 & 6.16 & 6.926 &  &  &  &   \\
DIT & 1 & 6 & 1.75 & 1.058 & 1-2 & 3-6 &  &   \\
NOC & 0 & 26 & .31 & 1.673 &  &  &  &   \\
CBO & 0 & 58 & 5.95 & 7.294 & 0-1 & 2-3 & 4-58 &   \\
RFC & 0 & 177 & 19.84 & 22.555 & 0-29 & 30-177 &  &   \\
LCOM & 0 & 2197 & 24.68 & 119.657 &  &  &  &   \\
Ca & 0 & 79 & 3.69 & 6.288 & 0-1 & 2-79 &  &   \\
NPM & 0 & 70 & 5.24 & 6.284 &  &  &  &   \\
\hline
\end{tabular}
\end{table}

\begin{table}[!t]
\renewcommand{\arraystretch}{1.3}
\caption{Descriptive Statistics for RapidMiner}
\label{table_metrics_RapidMiner}
\centering
\begin{tabular}{|c|c|c|c|c|c|c|c|c|c|}
\hline
 & Min & Max & Mean & Std. Deviation & Bin=1 & Bin=2 & Bin=3 & Bin=4 & Bin=5 \\
\hline
D-layer & 0 & 16 & 7.16 & 3.256  & 0-4 & 5-8 & 9-12 & 13-16 & \\
WMC & 0 & 163 & 5.81 & 7.909 & 0-1 & 2 & 3-10 & 11-163 &  \\
DIT & 1 & 11 & 2.68 & 1.878 & 1-2 & 3 & 4-7 & 8-11 & \\
NOC & 0 & 258 & .45 & 4.918 &  &  &  &  & \\
CBO & 0 & 121 & 6.44 & 8.368 & 0 & 1 & 2-3 & 4-14 & 15-121 \\
RFC & 0 & 456 & 21.25 & 27.964 & 0-2 & 3 & 4-27 & 28-107 & 108-456 \\
LCOM & 0 & 12887 & 31.42 & 305.142 & 0 & 1-4 & 5-35 & 36-12887 & \\
Ca & 0 & 767 & 6.17 & 32.375 & 0 & 1-2 & 3-27 & 28-767 & \\
NPM & 0 & 123 & 4.31 & 6.186 & 0 & 1 & 2 & 3-6 & 7-123 \\
\hline
\end{tabular}
\end{table}

\begin{table}[!t]
\renewcommand{\arraystretch}{1.3}
\caption{Descriptive Statistics for SweetHome3D}
\label{table_metrics_SweetHome3D}
\centering
\begin{tabular}{|c|c|c|c|c|c|c|c|c|}
\hline
 & Min & Max & Mean & Std. Deviation & Bin=1 & Bin=2 & Bin=3 & Bin=4\\
\hline
D-layer & 0 & 15 & 8.67 & 3.413 & 0-3 & 4-7 & 8-11 & 12-15 \\
WMC & 0 & 264 & 5.31 & 12.255 & 0-1 & 2 & 3-264 &  \\
DIT & 1 & 8 & 1.74 & 1.302 &  &  &  &  \\
NOC & 0 & 35 & .12 & 1.250 &  &  &  &  \\
CBO & 0 & 99 & 4.24 & 7.950 & 0-1 & 2-99 &  &  \\
RFC & 0 & 636 & 18.32 & 41.818 & 0-2 & 3-14 & 15-636 &  \\
LCOM & 0 & 28064 & 64.62 & 868.015 & 0 & 1-28064 &  &  \\
Ca & 0 & 271 & 4.17 & 12.737 & 0-2 & 3-271 &  &  \\
NPM & 0 & 77 & 3.04 & 6.332 & 0-1 & 2-77 &  &  \\
\hline
\end{tabular}
\end{table}

Next, Spearman correlation coefficient is calculated to reveal correlations among the metrics and the results are illustrated in Tables \ref{table_cor_JabRef}-\ref{table_cor_SweetHome3D}. While different design metrics appear to be correlated to D-layer, DIT, CBO, RFC, LCOM and Ca are always correlated to D-layer in all datasets.

D-layer and correlated variables are discretized to categorical variables. D-layer is binned to four categories, one per \emph{tentative} architecture layer. Variables are discretized using Minimal Description Length Principle (MDLP) discretization according to D-layer. Finally, classification rules are generated to predict the architecture layer of a class, D-layer. The discretized values are imported to Weka \cite{Hall2009} and JRip classification algorithm is applied. The classification rules generated for each project are shown in Appendix \ref{classification_rules}.

\begin{table}[!t]
\renewcommand{\arraystretch}{1.3}
\caption{Correlations for JabRef}
\label{table_cor_JabRef}
\centering
\tabcolsep 2.2pt
\begin{tabular}{|l|l|l|l|l|l|l|l|l|l|}
\hline
 & D-layer & WMC & DIT & NOC & CBO & RFC & LCOM & Ca & NPM \\
\hline
D-layer & 1,000 & -,113** & ,045* & -,117** & ,341** & ,078** & -,164** & -,141** & -,089** \\
WMC &  & 1,000 & ,032 & ,201** & ,448** & ,803** & ,680** & ,265** & ,857** \\
DIT &  &  & 1,000 & ,039* & -,017 & ,082** & ,015 & ,042* & ,014 \\
NOC &  &  &  & 1,000 & ,062** & ,132** & ,176** & ,279** & ,186** \\
CBO &  &  &  &  & 1,000 & ,677** & ,244** & ,056** & ,405** \\
RFC &  &  &  &  &  & 1,000 & ,479** & ,161** & ,684** \\
LCOM &  &  &  &  &  &  & 1,000 & ,194** & ,548** \\
Ca &  &  &  &  &  &  &  & 1,000 & ,268** \\
NPM &  &  &  &  &  &  &  &  & 1,000 \\
\hline
\end{tabular}
\end{table}

\begin{table}[!t]
\renewcommand{\arraystretch}{1.3}
\caption{Correlations for Jbpm}
\label{table_cor_Jbpm}
\centering
\tabcolsep 2.2pt
\begin{tabular}{|l|l|l|l|l|l|l|l|l|l|}
\hline
 & D-layer & WMC & DIT & NOC & CBO & RFC & LCOM & Ca & NPM \\
\hline
D-layer & 1,000 & ,014 & ,304** & ,001 & ,566** & ,238** & -,141** & -,231** & -,046** \\
WMC &  & 1,000 & ,132** & ,246** & ,301** & ,736** & ,698** & ,247** & ,925** \\
DIT &  &  & 1,000 & ,044 & ,319** & ,259** & ,282** & -,115** & ,051 \\
NOC &  &  &  & 1,000 & ,142** & ,212** & ,126** & ,300** & ,226** \\
CBO &  &  &  &  & 1,000 & ,679** & ,342** & -,188** & ,229** \\
RFC &  &  &  &  &  & 1,000 & ,551** & -,022 & ,636** \\
LCOM &  &  &  &  &  &  & 1,000 & ,088* & ,602** \\
Ca &  &  &  &  &  &  &  & 1,000 & ,269** \\
NPM &  &  &  &  &  &  &  &  & 1,000 \\
\hline
\end{tabular}
\end{table}

\begin{table}[!t]
\renewcommand{\arraystretch}{1.3}
\caption{Correlations for RapidMiner}
\label{table_cor_RapidMiner}
\centering
\tabcolsep 2.2pt
\begin{tabular}{|l|l|l|l|l|l|l|l|l|l|}
\hline
 & D-layer & WMC & DIT & NOC & CBO & RFC & LCOM & Ca & NPM \\
\hline
D-layer & 1000 & ,023 & ,228** & -,011 & ,427** & ,159** & ,062** & -,276** & ,007 \\
WMC &  & 1000 & ,181** & ,231** & ,414** & ,761** & ,750** & ,271** & ,892** \\
DIT &  &  & 1000 & ,069** & ,466** & ,347** & ,184** & -,116** & ,108** \\
NOC &  &  &  & 1000 & ,121** & ,157** & ,234** & ,352** & ,213** \\
CBO &  &  &  &  & 1000 & ,710** & ,367** & -,046** & ,329** \\
RFC &  &  &  &  &  & 1000 & ,539** & ,127** & ,651** \\
LCOM &  &  &  &  &  &  & 1000 & ,173** & ,685** \\
Ca &  &  &  &  &  &  &  & 1000 & ,234** \\
NPM &  &  &  &  &  &  &  &  & 1000 \\
\hline
\end{tabular}
\end{table}

\begin{table}[!t]
\renewcommand{\arraystretch}{1.3}
\caption{Correlations for SweetHome3D}
\label{table_cor_SweetHome3D}
\centering
\tabcolsep 2.2pt
\begin{tabular}{|l|l|l|l|l|l|l|l|l|l|}
\hline
 & D-layer & WMC & DIT & NOC & CBO & RFC & LCOM & Ca & NPM \\
\hline
D-layer & 1,000 & -,351** & -,087** & -,148** & ,193** & -,159** & -,345** & -,340** & -,410** \\
WMC &  & 1,000 & ,350** & ,206** & ,244** & ,699** & ,736** & ,489** & ,896** \\
DIT &  &  & 1,000 & ,051 & ,086** & ,231** & ,346** & ,216** & ,286** \\
NOC &  &  &  & 1,000 & ,035 & ,175** & ,176** & ,313** & ,176** \\
CBO &  &  &  &  & 1,000 & ,530** & ,079** & ,068* & ,203** \\
RFC &  &  &  &  &  & 1,000 & ,437** & ,321** & ,628** \\
LCOM &  &  &  &  &  &  & 1,000 & ,562** & ,634** \\
Ca &  &  &  &  &  &  &  & 1,000 & ,447** \\
NPM &  &  &  &  &  &  &  &  & 1,000 \\
\hline
\end{tabular}
\end{table}

The generated rules can partially reveal classes' behavior in respect to the architecture layer. For User Interface classes, D-layer 4, rules state that they must have low LCOM and low/medium DIT. We would suspect that User Interface classes should have such attributes since their main purpose is to interact with the system's user. For D-layer 3, JRip produced rules that consider Controller classes should have relatively low CBO and Ca. This assumption is expected since Controller classes are responsible for obtaining business logic data and passing them to User Interface classes. We would expect that Business Logic classes should have high DIT, CBO and WMC. JRip rules partially confirm our hypothesis, since rules for D-layer 2 include relatively high CBo and WMC. Finally, Infrastructure classes' rules show relatively low CBO and RFC since their main purpose is data storage.

In terms of precision and recall, Table \ref{table_accuracy_measures} shows the results for each project. CK metrics and JRip classification rules achieve attribute representation according to architecture layer, but they cannot form a solid classification criterion. For example, D-layer 3 and 4 classification rules for Jbpm and D-layer 4 for SweetHome3D are not able to recover layers' classes. The reason why higher architectural layers recovery cannot always be established, could be systems' \emph{tentative} architectural decomposition.

\begin{table}[!t]
\renewcommand{\arraystretch}{1.3}
\caption{Accuracy Measures}
\label{table_accuracy_measures}
\centering
\tabcolsep 2.2pt
\begin{tabular}{|c|c|c|c|c|c|c|c|c|}
\hline
 & \multicolumn{2}{|c|}{JabRef} & \multicolumn{2}{|c|}{Jbpm} & \multicolumn{2}{|c|}{RapidMiner} & \multicolumn{2}{|c|}{SweetHome3D} \\
 \hline
 & Precision & Recall & Precision & Recall & Precision & Recall & Precision & Recall  \\
\hline
D-layer=1 & 0.518 & 0.877 & 0.653 & 0.845 & 0.727 & 0.5 & 0.582 & 0.194 \\
\hline
D-layer=2 & 0.527 & 0.111 & 0.705 & 0.845 & 0.666 & 0.886 & 1 & 0.009 \\
\hline
D-layer=3 & 0.609 & 0.047 & 0 & 0 & 0.6 & 0.374 & 0.683 & 0.981 \\
\hline
D-layer=4 & 0.748 & 0.607 & 0 & 0 & 0.421 & 0.043 & 0 & 0 \\
\hline
\end{tabular}
\end{table}

\section{Conclusion and Future Research Directions}
\label{sec_conclusion}
This work aims to verify that object-oriented metrics can reveal architectural layer information. Our methodology is tested on four large open source software systems to recover their architectural layers. The results confirmed that several design metrics can assist in systems' architectural recovery. More precisely, DIT, CBO, RFC, LCOM and Ca appear to be correlated to the architecture layer in our datasets. However, our methodology requires further improvement to capture classes' behavior in all architectural layers.

Concluding, this work verified our assumption that design metrics can assist architecture layer recovery. Our following research includes further improvements, such as weight attribution to metric contribution and more metrics that could assist architecture recovery.

\subsubsection*{Acknowledgments.} This work is partially funded by the European Commission in the context of the OPEN-SME ``Open-Source Software Reuse Services for SMEs'' project, under the grant agreement no. FP7-SME-2008-2/243768.

\appendix
\section{Classification Rules}
\label{classification_rules}
\emph{JabRef Classification Rules}
\newline \emph{1.} IF (CBOBin = 4) and (NPMBin = 5) and (WMCBin = 5) and (CaBin = 1) THEN layerBin=3
\newline \emph{2.} IF (CBOBin = 4) and (WMCBin = 5) and (CaBin = 2) THEN layerBin=3
\newline \emph{3.} IF (CaBin = 1) and (WMCBin = 4) and (CBOBin = 3) and (RFCBin = 4) THEN layerBin=2
\newline \emph{4.} IF (WMCBin = 4) and (CaBin = 1) and (CBOBin = 3) and (NPMBin = 5) and (LCOMBin = 2) THEN layerBin=2
\newline \emph{5.} IF (WMCBin = 4) and (CBOBin = 3) and (CaBin = 1) and (NPMBin = 4) THEN layerBin=2
\newline \emph{6.} IF (WMCBin = 4) and (CBOBin = 3) and (NPMBin = 1) and (RFCBin = 5) and (CaBin = 2) THEN layerBin=2
\newline \emph{7.} IF (WMCBin = 3) and (LCOMBin = 1) and (NPMBin = 2) THEN layerBin=4
\newline \emph{8.} IF (CBOBin = 4) and (DITBin = 2) THEN layerBin=4
\newline \emph{9.} IF (WMCBin = 3) and (LCOMBin = 1) and (CBOBin = 3) and (NPMBin = 3) THEN layerBin=4
\newline \emph{10.} IF (CBOBin = 4) and (NOCBin = 1) and (LCOMBin = 1) and (NPMBin = 3) THEN layerBin=4
\newline \emph{11.} IF (NPMBin = 4) and (CBOBin = 3) and (LCOMBin = 2) and (NOCBin = 1) and (CaBin = 2) THEN layerBin=4
\newline \emph{12.} IF (RFCBin = 6) and (NPMBin = 4) and (LCOMBin = 1) THEN layerBin=4
\newline \emph{13.} IF (CBOBin = 4) and (NOCBin = 1) and (RFCBin = 6) and (WMCBin = 4) and (CaBin = 3) and (NPMBin = 4) THEN layerBin=4
\newline \emph{14.} IF (WMCBin = 3) and (CaBin = 2) and (CBOBin = 2) and (LCOMBin = 2) THEN layerBin=4
\newline \emph{15.} IF (CBOBin = 4) and (NOCBin = 1) and (RFCBin = 6) and (WMCBin = 4) and (NPMBin = 5) and (CaBin = 1) THEN layerBin=4
\newline \emph{16.} ELSE layerBin=1
\newline \emph{Jbpm Classification Rules}
\newline \emph{1.} IF (DITBin = 2) and (CBOBin = 3)THEN layerBin=2
\newline \emph{2.} IF (CBOBin = 3) and (RFCBin = 2) THEN layerBin=2
\newline \emph{3.} IF (DITBin = 2) and (CBOBin = 2) THEN layerBin=2
\newline \emph{4.} ELSE layerBin=1
\newline \emph{RapidMiner Classification Rules}
\newline \emph{1.} IF (CBOBin = 5) and (CaBin = 2) and (DITBin = 2) and (RFCBin = 4) and (LCOMBin = 1) THEN layerBin=4
\newline \emph{2.} IF (CBOBin = 1) and (RFCBin = 3) THEN layerBin=1
\newline \emph{3.} IF (CBOBin = 2) and (CaBin = 3) and (RFCBin = 3) and (DITBin = 1) THEN layerBin=1
\newline \emph{4.} IF (CBOBin = 2) and (LCOMBin = 2) and (CaBin = 1) and (DITBin = 1) THEN layerBin=1
\newline \emph{5.} IF (DITBin = 1) and (CBOBin = 1) and (CaBin = 1) THEN layerBin=1
\newline \emph{6.} IF (DITBin = 1) and (CBOBin = 1) and (CaBin = 3) THEN layerBin=1
\newline \emph{7.} IF (CBOBin = 2) and (CaBin = 4) THEN layerBin=1
\newline \emph{8.} IF (CaBin = 1) and (LCOMBin = 2) THEN layerBin=3
\newline \emph{9.} IF (LCOMBin = 3) and (CaBin = 1) THEN layerBin=3
\newline \emph{10.} IF (DITBin = 2) and (LCOMBin = 1) and (CBOBin = 5) THEN layerBin=3
\newline \emph{11.} IF (DITBin = 2) and (CaBin = 2) and (CBOBin = 3) and (LCOMBin=1) THEN layerBin=3
\newline \emph{12.} IF (CaBin = 2) and (RFCBin = 4) and (CBOBin = 5) THEN layerBin=3
\newline \emph{13.} ELSE layerBin=2
\newline \emph{SweetHome3D Classification Rules}
\newline \emph{1.} IF (CBOBin = 1) and (CaBin = 2) and (RFCBin = 1) THEN layerBin=1
\newline \emph{2.} IF (CaBin = 2) and (CBOBin = 1) and (NPMBin = 2) and (WMCBin = 2) THEN layerBin=1
\newline \emph{3.} ELSE layerBin=3
\end{document}